\documentclass[a4paper,11pt]{article}
\usepackage{pos}
\usepackage{booktabs}

\newcommand{\ce}[1]{Eq.~(\ref{#1})}
\newcommand{\cf}[1]{{Fig.~\ref{#1}}}
\newcommand{\ct}[1]{{Table~\ref{#1}}}

\title{Updates on transversity extractions}
\author*[a,b]{Carlo Flore}

\affiliation[a]{Dipartimento di Fisica, Università di Cagliari, Cittadella Universitaria, I-09042 Monserrato (CA), Italy}
\affiliation[b]{INFN, Sezione di Cagliari, Cittadella Universitaria, I-09042 Monserrato (CA), Italy}

\emailAdd{carlo.flore@unica.it}

\abstract{We give a brief overview on some of the latest results on the transversity function. We focus on some recent extractions from azimuthal asymmetries data measured in semi-inclusive deep-inelastic scattering processes, and on the impact of transverse single-spin asymmetry data measured in polarised proton-proton collisions on these extractions.}

\FullConference{7th International Workshop on “Transverse phenomena in hard processes and the transverse structure of the proton (Transversity2024)\\
 3-7 June 2024\\
Trieste, Italy\\}

\tableofcontents

\begin{document}
\maketitle

\section{Introduction}

The transversity function, $h_1^q (x)$, is one of the three independent functions describing the collinear structure of spin-$\tfrac{1}{2}$ hadrons at leading twist. Being a chiral-odd quantity, it is not accessible in deep-inelastic scattering (DIS) processes. Thus, it has to be coupled with another chiral-odd quantity. In the context of collinear perturbative QCD, $h_1^q(x)$ is accessed together with di-hadron fragmentation functions (FFs) in two-hadron production in proton-proton and lepton-proton collisions~\cite{Bacchetta:2012ty, Martin:2014wua, Radici:2018iag, Benel:2019mcq}, or in the framework of transverse momentum dependent distributions (TMDs), together with the Collins FF in semi-inclusive DIS (SIDIS) processes~\cite{Anselmino:2007fs, Anselmino:2013vqa, DAlesio:2020vtw}. 

$h_1^q$ is related to the two other independent collinear distributions (unpolarised and helicity PDFs) by the bound derived by Soffer~\cite{Soffer:1994ww}:
  \begin{equation}
   \lvert h_1^q(x, Q^2) \rvert \leq \frac{1}{2} \left[ f_{q/p}(x, Q^2) + g_{1L}^q(x, Q^2)\right] \equiv {\rm SB}^q(x, Q^2)\,.
  \end{equation}
The Soffer bound (SB) was shown to be preserved by $Q^2$ evolution up to next-to-leading order in QCD~\cite{Barone:1997fh,Vogelsang:1997ak}, and it represents a useful constraint for phenomenological analyses.

The interest in transversity extractions goes beyond the description of hadron structure.  
Indeed, quarks contribute to the nucleon tensor charge via the first Mellin moment of the non-singlet quark combination, defined as:
 \begin{equation}
 \delta q = \int_0^1\,\left[h_1^{q}(x) - h_1^{\bar q}(x)\right]\,dx\,,
\label{eq:delta-q}
 \end{equation}
and the isovector combination of tensor charges 
 \begin{equation}
 g_T = \delta u - \delta d
 \label{eq:g_T}
\end{equation}
represents also an interesting quantity for beyond Standard Model (BSM) effects~\cite{Cirigliano:2009wk,Bhattacharya:2011qm,Courtoy:2015haa}. $\delta q$ and $g_T$ are also intensively studied within lattice QCD~\cite{Alexandrou:2024awx}. Therefore, transversity-related studies represent a bridge between QCD phenomenology, lattice QCD and BSM physics.

Here, we will concentrate on the latest results on transversity extractions within the TMD framework, touching upon different issues such as the usage of the SB and the compatibility of $h_1^q$ extractions with complementary data from $p^\uparrow p$ collisions.

\section{Latest results from SIDIS data}

We start by summarising the results of Ref.~\cite{DAlesio:2020vtw}, where the issue of the usage of the SB in the transversity extraction was thoroughly investigated. When extracting the transversity function, it is customary to adopt, at the initial scale $Q_0^2$, a parametrisation proportional to the SB~\cite{Anselmino:2007fs,Anselmino:2013vqa,Bacchetta:2012ty,Radici:2018iag}
\begin{equation}
h_1^q(x, Q_0^2) \propto {\rm SB}(x,Q_0^2)\,.
\end{equation}
The functional forms are written in a way such that the SB is automatically fulfilled for every $x$ and $Q^2$ values throughout the fit. This choice represents a potential extra bias for the extraction: the amount of data available for the fit is not always large enough, and is usually not covering a sufficiently wide $x$-region. Therefore, when computing $\delta q$ and $g_T$, their value mostly results from an extrapolation that depends on the selected functional form for $h_1^q$. 

In Ref.~\cite{DAlesio:2020vtw} we proposed to avoid the automatic fulfillment of the SB in the parametrisation, but to apply it {\it a posteriori} on the extracted transversity functions. To illustrate the new method, we updated the extraction of Ref.~\cite{Anselmino:2015sxa}, where the transversity function is parametrised as:
\begin{align}
& h_1^q(x, Q^2, k^2_\perp) = h_1^q (x, Q^2) \frac{e^{-k^2_\perp / \langle k^2_\perp \rangle}}{\pi \langle k^2_\perp\rangle}\,,\\  \nonumber
h_1^q(x, Q_0^2) &= N^{T}_q x^{\alpha} (1-x)^\beta \,
\frac{(\alpha+\beta)^{\alpha+\beta}}{\alpha^\alpha \beta^\beta}
{\rm SB}^q(x, Q_0^2)
 \end{align}
for $q = u_v,\,d_v$. Upon constraining $\lvert N_q^T \rvert \leq 1$ the SB is automatically fulfilled. Within the new approach, such a constraint is no longer adopted on the parametrisation, but rather imposed on the Monte Carlo (MC) sets generated for estimating the uncertainty on $h_1^q$. In doing so, we removed this extra bias, and we are able to check if the extracted transversity PDFs are compatible with the SB. 

The results are presented in \cf{fig:h1-usingSB-vs-noSB}, where we dubbed as ``using SB'' and ``no SB'' respectively the cases in which we apply the SB a posteriori and the one in which the SB is not applied at all. We note that: (a) the two extractions have almost the same $\chi^2_{\rm dof} \approx 0.93$; (b) the application of the SB a posteriori allows to properly estimate the size of the $d_v$ transversity function and its uncertainty (cfr.~{\it e.g.}~Fig.~7 of Ref.~\cite{Anselmino:2015sxa}); (c) when relaxing the SB constraint, while the extracted $h_1^{u_v}(x)$ does not change very much, $h_1^{d_v}$ apparently violates the SB; (d) the violation has a statistical significance smaller than $1 \sigma$ where data is available (white background in the plots of \cf{fig:h1-usingSB-vs-noSB}).

\begin{figure}[h!]
\centering
\includegraphics[width=7cm, draft=false, keepaspectratio]{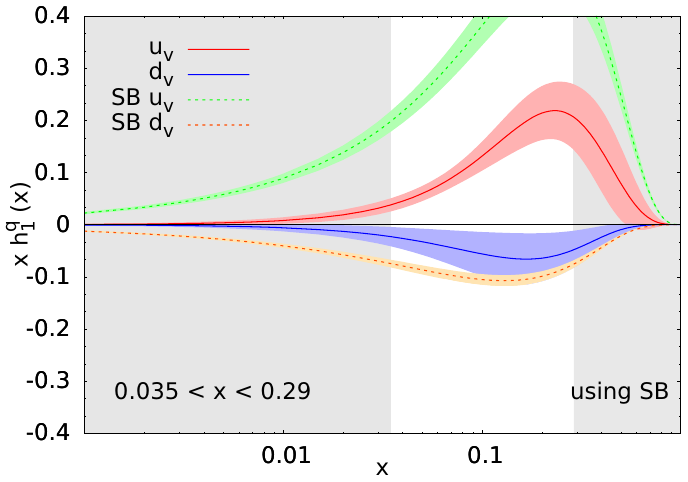}
\includegraphics[width=7cm, draft=false, keepaspectratio]{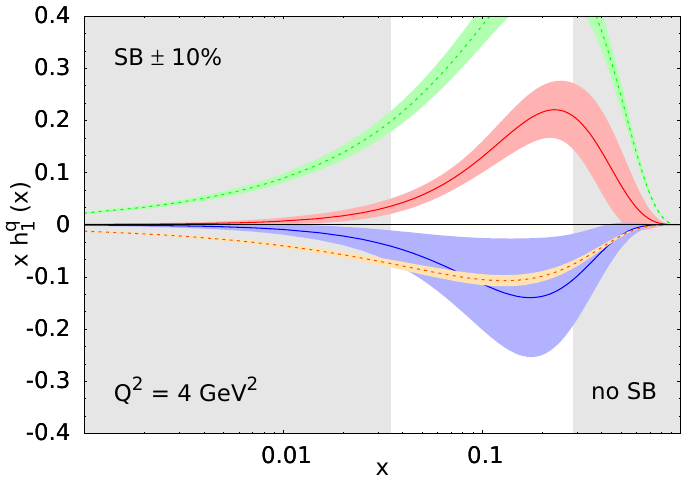}
\caption{Comparison of extracted transversity functions for $u_v$ and $d_v$ with the application of the SB a posteriori (left) or without applying the SB (right). Figure taken from Ref.~\cite{DAlesio:2020vtw}.}
\label{fig:h1-usingSB-vs-noSB}
\end{figure}

%The application of the SB a posteriori allows also to ease the tension with lattice QCD estimates of $\delta q$ and $g_T$. We 

Later, in Ref.~\cite{Boglione:2024dal}, we updated again the extraction of the transversity functions by including the latest data from the HERMES Collaboration~\cite{HERMES:2020ifk}. A comparison of the two extractions of Ref.~\cite{DAlesio:2020vtw} (dubbed as ``fit 2020'', ``using SB'' case) and of Ref.~\cite{Boglione:2024dal} (``fit 2023'') is presented in \cf{fig:h1-fit20vs23}. Note that the two extractions adopted a different collinear PDF set, namely the CTEQ66~\cite{Nadolsky:2008zw} and the MSHT20nlo~\cite{Bailey:2020ooq} sets, respectively. On the other hand, we used the same collinear helicity PDF set from DSSV~\cite{deFlorian:2009vb} and the same collinear FFs for pions and kaons from DEHSS~\cite{deFlorian:2014xna, deFlorian:2017lwf}\footnote{We guide the reader to~\cite{DAlesio:2020vtw, Boglione:2024dal} for all the details about the adopted parametrisations.}. The main difference between the two extraction is in the magnitude of $h_1^{u_v}$, whose corresponding normalization value is, on average, larger than the one of the previous extraction, as shown in the right panel of \cf{fig:h1-fit20vs23}. Nevertheless, the two extractions are compatible with each other. In the future, the new  COMPASS measurements~\cite{COMPASS:2023vhr} are expected to further reduce the uncertainties on the extracted transversity functions.

\begin{figure}[htbp]
\centering
\includegraphics[width=6.8cm,draft=false]{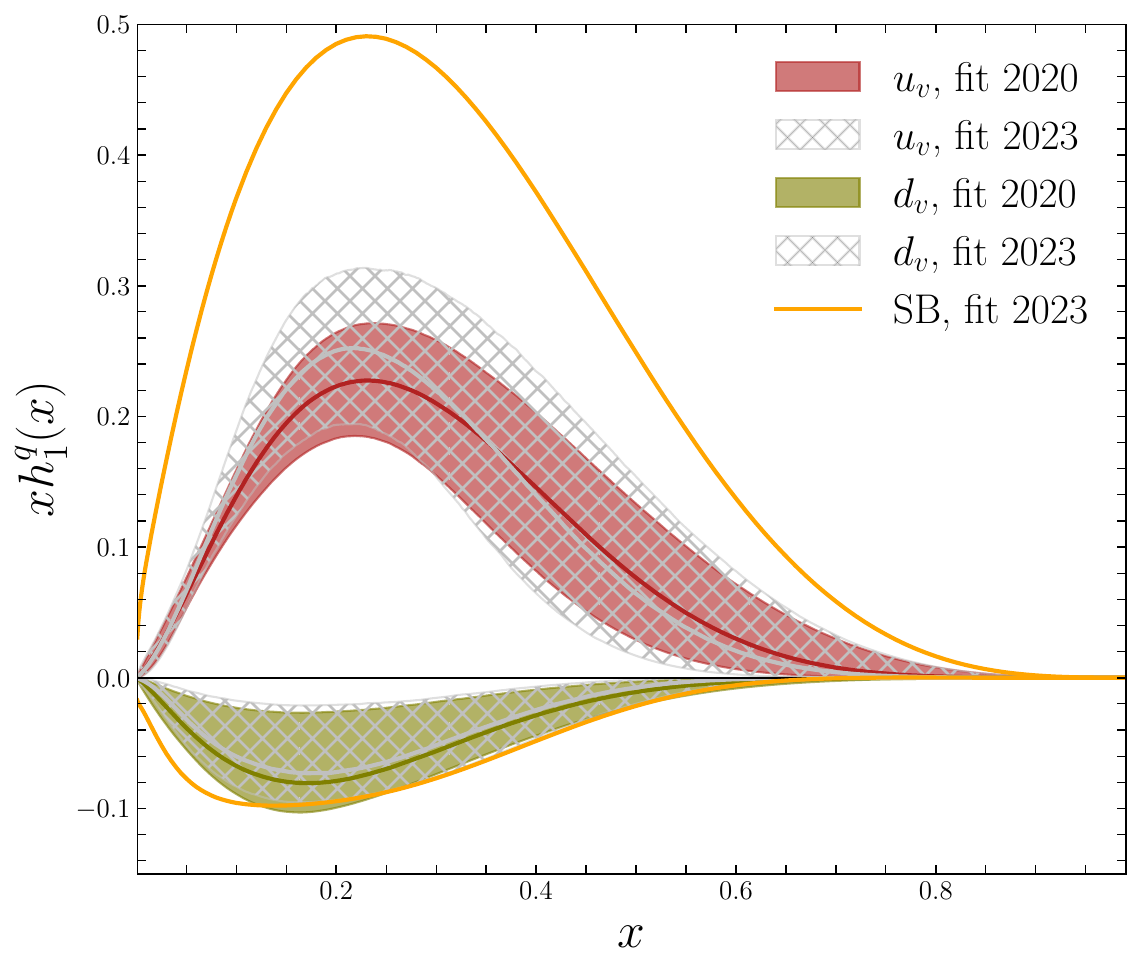}
\includegraphics[width=7cm,draft=false]{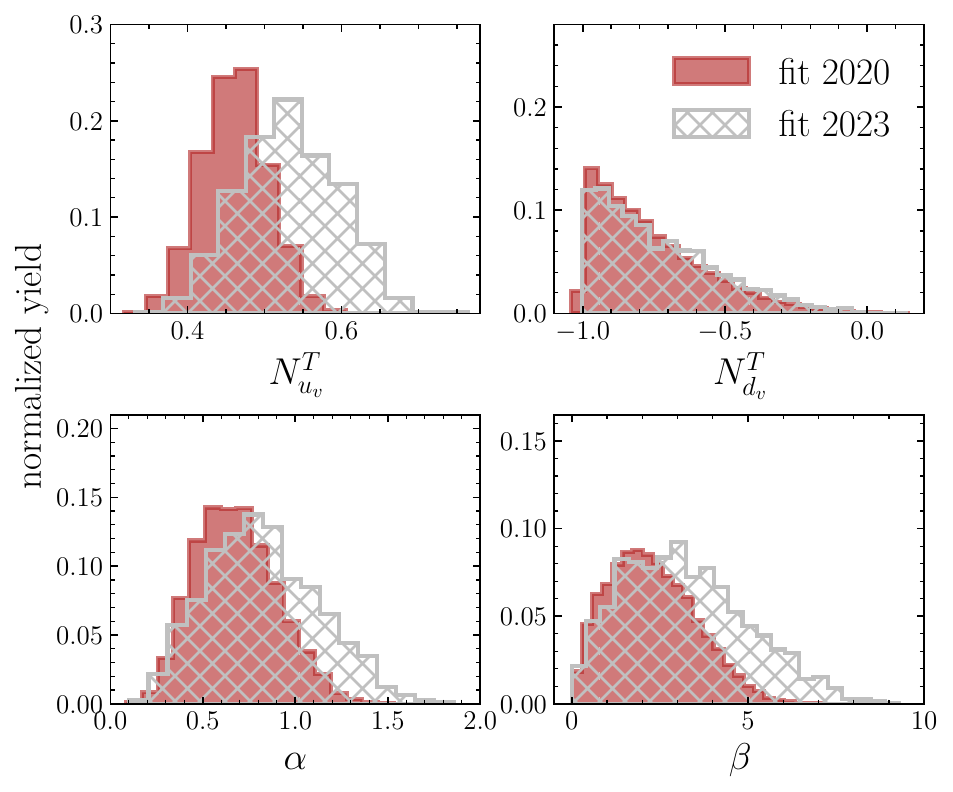}
\caption{Left: comparison of extracted transversity functions from Ref.~\cite{DAlesio:2020vtw} (``fit 2020'') and Ref.~\cite{Boglione:2024dal} (``fit 2023'') and corresponding Soffer bound for the extraction of Ref.~\cite{Boglione:2024dal} . Right: comparison of parameter distributions for the two extractions.}
\label{fig:h1-fit20vs23}
\end{figure}

\section{Impact of $A_N$ data}

As previously mentioned, SIDIS data are limited in their kinematical coverage. Hence, complementary data are needed to reduce the extent to which the extrapolation for $\delta q$ and $g_T$ is performed. 

Another proxy to the transversity function are the transverse single-spin asymmetries (TSSA or $A_N$) measured in $p^\uparrow p \to h X$ processes. These reactions can be described within the Generalised Parton Model (GPM)~\cite{Anselmino:2005sh}, where a factorised formulation in terms of TMDs is assumed as a starting point for the cross section, or within the Colour Gauge Invariant extension of the GPM (CGI-GPM)~\cite{Gamberg:2010tj}\footnote{This extension allows to recover the Sivers sign change through a one gluon exchange approximation.}. 

The TSSA is defined as:
\begin{equation}
   A_N = \frac{d\sigma^\uparrow-d\sigma^\downarrow}{d\sigma^\uparrow+d\sigma^\downarrow} =\, \frac{d\Delta\sigma}{ 2 d\sigma} \simeq \frac{d\Delta\sigma_{\text{Siv}} + d\Delta\sigma_{\text{Col}}}{ 2 d\sigma}\,,
   \label{eq:A_N}
  \end{equation}
where $d\sigma^\uparrow(\downarrow)$ is the polarised cross section for upward (downward) proton transverse polarisation, and where in the last equality we explicitly assume the (CGI-)GPM. The two terms at the numerator of \ce{eq:A_N} are related to the Sivers and Collins effects respectively. The latter is proportional to the convolution of the TMD transversity and the Collins FF:
\begin{equation}
 d\Delta\sigma_{\text{Col}} \propto \sum_{a,b,c,d} h_1^a(x_a, k_{\perp a}) \otimes f_{b/p}(x_b, k_{\perp b}) \otimes d\Delta\sigma^{a^\uparrow b \to c^\uparrow d} \otimes H_1^{\perp c}(z, k_{\perp h})\,,
\end{equation}
and is sensitive to the large-$x$ behaviour of $h_1^q$.

In Ref.~\cite{Boglione:2024dal} a Bayesian simultaneous reweighting was applied on the Sivers, transversity and Collins extractions from SIDIS and $e^+e^-$ data, using $A_N$ data measured by the BRAHMS and STAR Collaborations at RHIC. The results for the reweighted transversity functions and parameter distributions are presented in \cf{fig:h1-fit23-rew}. Some comments are in order: (a) $A_N$ data mostly impact on the transversity distribution; (b) the reweighted transversity distributions in the (CGI-)GPM formalism follow the SB shape rather closely at large $x$ (see also the $\alpha$ and $\beta$  distributions on the right panel of \cf{fig:h1-fit23-rew}); (c) the uncertainty reduction is up to 80-90\% for $h_1^q$ at large $x$; (d) the Collins mechanism turned out to be the dominant contribution to $A_N$\footnote{In the (CGI-)GPM this was never seen before applying the SB a posteriori on the MC sets, finding now consistency with the observations in the collinear twist-3 formalism~\cite{Gamberg:2022kdb}.}.

\begin{figure}[hbtp]
\centering
\includegraphics[width=6.9cm,draft=false]{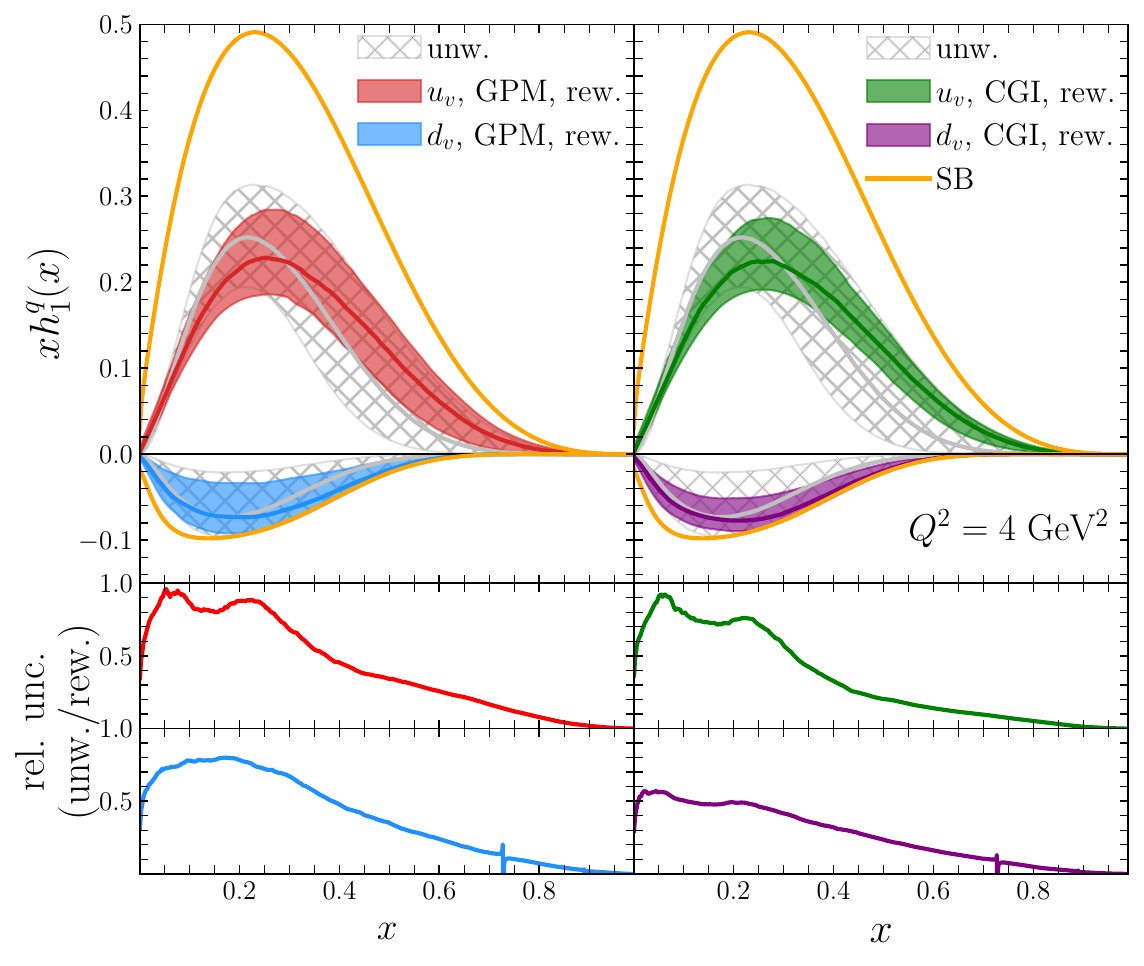}
\includegraphics[width=7cm,draft=false]{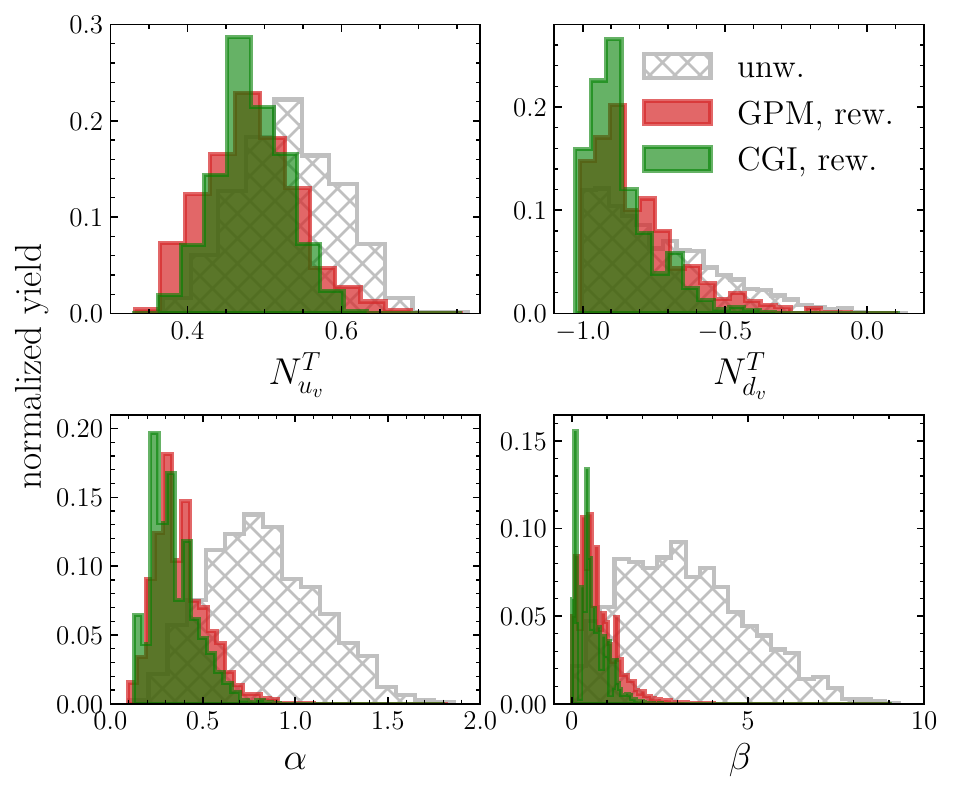}
\caption{Unweighted and reweighted transversity functions from Ref.~\cite{Boglione:2024dal} (left) and comparsion of unweighted and reweighted parameter distributions (right) in the GPM and the CGI-GPM models.}
\label{fig:h1-fit23-rew}
\end{figure}

\section{Tensor charges\label{sec:tensor-charges}}

Through \ce{eq:delta-q} and \ce{eq:g_T} we can compute the tensor charges at $Q^2 = 4$ GeV$^2$ for the two transversity extractions~\cite{DAlesio:2020vtw,Boglione:2024dal} we have presented here. We summarise the results below in \ct{tab:tensor-charges}.

\begin{table}[h!]
\centering
 \begin{tabular}{c c c c c c}
  \toprule
   \multicolumn{6}{c}{$Q^2 = 4$ GeV$^2$} \\
  \midrule
  ~ & \vtop{\hbox{\strut using SB}\hbox{\strut $\;$Ref.~\cite{DAlesio:2020vtw}}} & \vtop{\hbox{\strut $\;$no SB}\hbox{\strut Ref.~\cite{DAlesio:2020vtw}}}  & \vtop{\hbox{\strut $\;\;\,$unw.}\hbox{\strut Ref.~\cite{Boglione:2024dal}}}  & \vtop{\hbox{\strut GPM rew.}\hbox{\strut $\;$Ref.~\cite{Boglione:2024dal}}} & \vtop{\hbox{\strut CGI rew.}\hbox{\strut Ref.~\cite{Boglione:2024dal}}}\\
  \midrule
  $\delta u$ &  $0.42 \pm 0.09$ & $0.40 \pm 0.09$ & $0.46^{+0.10}_{-0.09}$ & $0.47^{+0.09}_{-0.07}$ & $0.47^{+0.08}_{-0.05}$ \\
  $\delta d$ & $-0.15 \pm 0.11$ & $-0.29 \pm 0.22$ & $-0.15^{+0.10}_{-0.07}$ & $-0.18^{+0.10}_{-0.06}$ & $-0.19^{+0.07}_{-0.05}$ \\
  $g_T$ & $0.57 \pm 0.13$ & $0.69 \pm 0.21$ & $0.60^{+0.13}_{-0.11}$ & $0.64^{+0.11}_{-0.09}$ & $0.65^{+0.10}_{-0.07}$\\
  \bottomrule
  \end{tabular}
  \caption{Tensor charges computed for the extractions of Ref.~\cite{DAlesio:2020vtw} and Ref.~\cite{Boglione:2024dal}, respectively with symmetric and asymmetric uncertainties at $2\sigma$ confidence level.}
  \label{tab:tensor-charges}
\end{table}

Finally, in \cf{fig:gT-comparison} we present a comparison of the results of Refs.~\cite{DAlesio:2020vtw,Boglione:2024dal} and various estimates of the tensor charges from phenomenological analyses. All of these analyses yield consistent values for $g_T$ , $\delta u$, and $\delta d$. This corroborates the consistency of different extractions of transversity within different approaches exploiting a variety of experimental data.

\begin{figure}[htbp]
\centering
\includegraphics[width=13cm,draft=false]{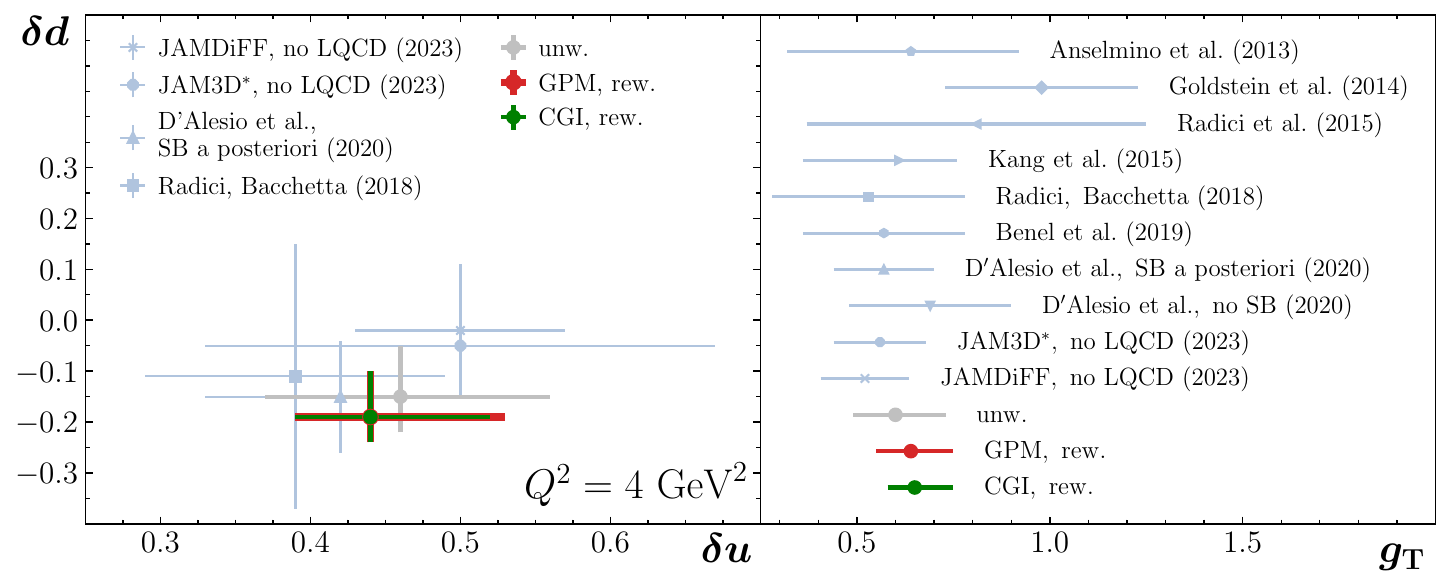}
\caption{Comparison of $u$ and $d$ tensor charges (left panel) and the iso-vector combination $g_T$ (right panel) from Ref.~\cite{Boglione:2024dal} with other phenomenological estimates at $Q^2= 4$ GeV$^2$. Figure taken from Ref.~\cite{Boglione:2024dal}. See references therein for the different results from other phenomenological extractions.}
\label{fig:gT-comparison}
\end{figure}

\section{Conclusions}

We have presented here the latest updates on transversity extractions within the TMD framework. We have studied the role of the Soffer bound in the determination of transversity and the tensor charges, proposing a new approach for the application of positivity bounds in phenomenological analyses. This procedure allows to properly explore the parameter space and to test whether theoretical expectations are met by experimental data. Furthermore, we have presented the results of a simultaneous Bayesian reweighting of the transversity function using $A_N$ data for polarised $pp$ scattering. $A_N$ data give further constraints on the large-$x$ behaviour of the transversity functions, and the corresponding tensor charge results corroborate the consistency of several extractions within different formalisms that probe $h_1^q $ in a variety of processes.

\acknowledgments
C.F.~is supported by the European Union ''Next Generation EU" program through the Italian PRIN 2022 grant n.~20225ZHA7W.

\bibliographystyle{JHEP}
\bibliography{references}

\end{document}